\documentclass[submission,copyright,creativecommons]{eptcs}
\providecommand{\event}{ESSS 2014} 

\usepackage[utf8]{inputenc}

\usepackage{graphicx}
\graphicspath{{images/}}

\usepackage{tikz}
\usetikzlibrary{arrows,petri,shadows,shapes}

\usepackage{caption}
\usepackage{subcaption}

\usepackage{listings}

\definecolor{mygreen}{rgb}{0,0.6,0}
\definecolor{mygray}{rgb}{0.5,0.5,0.5}
\definecolor{mymauve}{rgb}{0.58,0,0.82}

\newcommand{\paragraphe}[1]{\smallskip

\noindent\textbf{#1.}\ }

\newcommand{\CPNTools}{CPN\,Tools}
\newcommand{\OurTool}{UML2CPN}

\newcommand{\CDpPlaying}{\mbox{PL}}
\newcommand{\CDpPaused}{\mbox{PA}}

\newcommand{\PNnulltype}{\bullet}

\newcommand{\Pin}{p_{\mathit{in}}}
\newcommand{\Pout}{p_{\mathit{out}}}

	\newcommand{\vrai}{\mathtt{true}}
	\newcommand{\faux}{\mathtt{false}}

\ifdefined \VersionWithComments
	\usepackage{marginnote}
	
\newcommand{\marginX}{\marginnote{\huge{\quad\quad\textbf{!}\quad\quad}}}
	\newcommand{\commentaire}[1]{{\color{red}\marginX{}$\Leftarrow$ 
\textbf{#1} $\Rightarrow$}}
	\newcommand{\ea}[1]{}
	\newcommand{\gl}[1]{{\color{magenta}\marginX{}[\textbf{Giuseppe}: #1]}}
	\newcommand{\yc}[1]{{\color{green}\marginX{}[\textbf{Youcheng}: #1]}}
	
\else
	\newcommand{\commentaire}[1]{}
	\newcommand{\ea}[1]{}
	\newcommand{\gl}[1]{}
	\newcommand{\yc}[1]{}
\fi

\title{Translating UML State Machines to Coloured Petri Nets Using Acceleo: A Report}
\def\titlerunning{Translating UML State Machines to Coloured Petri Nets Using Acceleo: A Report}

\author{Étienne André, Mohamed Mahdi Benmoussa, Christine Choppy
\institute{Université Paris 13, Sorbonne Paris Cité, LIPN, CNRS, UMR 7030, F-93430, Villetaneuse, France}
}
\def\authorrunning{É. André, M.\,M. Benmoussa \& C. Choppy}
\begin{document}
\maketitle

\ifdefined \VersionWithComments
	\textcolor{red}{\textbf{To disable comments, comment out line~3 in the \LaTeX{} source.}}
\fi

\begin{abstract}
	UML state machines are widely used to specify dynamic systems behaviours.
	However its semantics is described informally, thus preventing the application of model checking techniques that could guarantee the system safety.
	In a former work, we proposed a formalisation of non-concurrent UML state machines using coloured Petri nets, so as to allow for formal verification.
	In this paper, we report our experience to implement this translation in an automated manner using the model-to-text transformation tool Acceleo.
	Whereas Acceleo provides interesting features that facilitated our translation process, it also suffers from limitations uneasy to overcome.
\end{abstract}


\ea{note: the main spelling seems to be ``metamodel'' with no space nor ``-''}

\section{Introduction}

UML~\cite{UML241} became the \emph{de facto} standard for modelling systems, 
and features a very rich syntax.
UML behavioural state machine diagrams (SMDs) are transition systems used to express the behaviour of dynamic systems in response to external interactions.
Although UML is widely used in the industry, its semantics is not formally expressed, 
hence not directly suitable for applying formal methods guaranteeing the system safety. 
However, a formal semantics can be given, for instance by translation to a formalism, 
such as coloured Petri nets (CPNs)~\cite{JK09}.
CPNs offer a detailed view of the process with a graphical representation.
One could argue that an intermediate formalism (as it is the case in this work) does not need to be clear or graphical.
However, we do believe that it is important and useful to provide a readable presentation that we can check against what we expect, and that could also suggest some improvements or some properties to check.
CPNs also benefit from powerful tools (such as \CPNTools{}~\cite{Westergaard13}) to test and check the model.
Translating an SMD to a CPN will allow us to formally guarantee the system safety by verifying it against properties.

In a previous work~\cite{ACK12}, we described a translation of SMDs into CPNs. 
These SMDs are non-concurrent, in the sense that synchronisation of events, fork and join pseudostates, are discarded.
However, we do consider history pseudostates (that remember where to go back, e.g.\ in case of failure), do/entry/exit behaviours, hierarchy of machines with inter-level transitions, and variables appearing in guards and behaviours.

\paragraphe{Objectives}
The translation introduced in~\cite{ACK12} was presented in an algorithmic form, but no implementation allowing an automated translation was performed, and that work did not contain precise indications to actually implement the translation.
In this paper, we present an attempt to perform an automated translation of SMDs into CPNs (following the transformation rules of~\cite{ACK12}) using model transformation techniques.
The lack of formal metamodel for CPNs together with the difficulty to build one drove us to use model-to-text transformation techniques, and we chose to use the model-to-text transformation tool Acceleo\footnote{\url{http://www.eclipse.org/acceleo/}}.
Whereas Acceleo provides interesting features that eased our translation process, it also suffers from limitations that made our translation not entirely straightforward. 
We report here our experience with Acceleo, and we point out both its advantages and limitations, 
while suggesting some new features that would make it more powerful.

\paragraphe{Related Work}
Verification of SMDs has been often tackled (see, e.g.~\cite{BR04} for a survey).
Some approaches directly give UML a semantics (e.g.~\cite{LLACSWD13}).
Many approaches translate UML specifications into an intermediate model of some model checker, e.g., SMV~\cite{McMillan92}
or SPIN~\cite{Holzmann03}. 
Other approaches (e.g.~\cite{PettitG06,LianHS08,CKZ10}) use CPNs as an intermediate formal model,
and use \CPNTools{} to analyse the generated CPN. 
The translation of~\cite{CKZ10}, where a formalisation of SMDs using CPNs is proposed, is probably the closest to
the technique presented in~\cite{ACK12}, although the sets of UML syntactic features taken into account are different.\ea{implemented how?}
%
We are not aware of previous attempts of translations of UML state machines to (extensions of) Petri nets using model transformation techniques.

\section{Preliminaries}

\begin{figure}[ht]
{\centering
	\begin{subfigure}[b]{0.62\textwidth}
		\includegraphics[width=\textwidth]{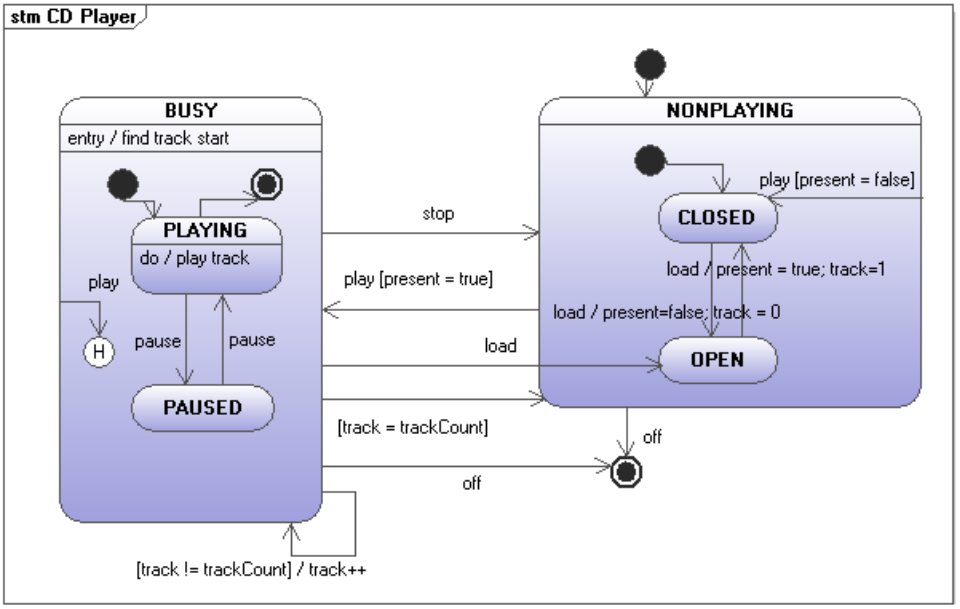}
		\caption{Specification using an SMD}
		\label{figure:CD:UML}
	\end{subfigure}

	\begin{subfigure}[b]{0.65\textwidth}
		\scalebox{.66}{
		\begin{tikzpicture}[scale=.7,node distance=1cm,>=stealth,auto]
			\tikzstyle{place}=[circle,thin,draw=blue!50!black,fill=blue!8,minimum size=16, inner sep=1, outer sep = 0]
			\tikzstyle{transition}=[rectangle,thin,draw=black!80, fill=gray!20,minimum size=6, inner sep=2, outer sep = 0]
			\tikzstyle{arc}=[->,semithick, color=black, draw=black]
			\tikzstyle{new}=[very thick, fill=purple!30, draw=purple!60!black, inner sep = 3]

			\tikzstyle{every label}=[blue!50!black]
			
			\node [place] at (0, 18.5) (Pin) {$\Pin$};
			\node [transition] at (0, 17.25) (TFTS) {$\mathit{FTS}$};
			\node [place] at (0, 16) (PFTS) {FTS};
			\node [transition] at (+1, 15) (T2) {};
			\node [transition] at (-1, 15) (T2') {};
			\node [transition] at (+3, 15) (TSL) [label=right:{$[t \neq N]$}] {};
			\node [transition] at (-3, 15) (TSL') [label=left:{$[t \neq N]$}] {};
			\node [place] at (+2, 16) (P2) {};
			\node [place] at (-2, 16) (P2') {};
			\node [place] at (0, 12) (Playing) {$\CDpPlaying$};
			\node [transition] at (3, 12) (PT) {$\mathit{PT}$};
			\node [transition] at (3, 14) (Playing-BUSYF) {};
			\node [place] at (5, 14) (BUSYF) {Busy$^F$};
			\node [transition] at (-2, 13) (H-Playing)
		[label=left:{$[h=\CDpPlaying]$}] {};
			\node [transition] at (-2, 12) (Playing-H) {play};
			\node [place] at (-4, 10) (BusyH) {Busy$^H$};
			\node [transition] at (+1, 10) (Playing-Paused) {pause};
			\node [transition] at (-1, 10) (Paused-Playing) {pause};
			\node [transition] at (-2, 8) (H-Paused) [label=left:{$[h=\CDpPaused]$}]
		{};
			\node [transition] at (-2, 9) (Paused-H) {play};
			\node [place] at (0, 8) (Paused) {$\CDpPaused$};
			
			\path
				(Pin) edge [arc] node {$vh$} (TFTS)
				(TFTS) edge [arc] node {$vh$} (PFTS)
				
				(Playing) edge [arc, above] node [xshift=5, yshift=10] {$vh$}
		(TSL)
				(TSL) edge [arc, bend right, above right] node {$p,t+1,h$} (Pin)
				(TSL) edge [arc, bend right] node {$\PNnulltype$} (P2)
				(PFTS) edge [arc, bend left, below left] node {$vh$} (T2)
				(P2) edge [arc, bend right] node {$\PNnulltype$} (T2)
				(T2) edge [arc, bend angle=20, bend left, above left] node
		{$p,t,\CDpPlaying$} (Playing)
				(Paused) [arc] --++ (0, -1) --++ (-5.05, 0) node [above right]
		{$vh$} --++ (0, 7) -- (TSL')
				(TSL') edge [arc, bend left] node {$p,t+1,h$} (Pin)
				(TSL') edge [arc, bend left, below right] node {$\PNnulltype$}
		(P2')
				(PFTS) edge [arc, bend right] node {$vh$} (T2')
				(P2') edge [arc, bend left, below left] node {$\PNnulltype$}
		(T2')
				(T2') edge [arc, bend angle=20, bend right, above left] node
		{$p,t,\CDpPlaying$} (Playing)

				(Playing) edge [arc, bend left, below] node {$vh$} (PT)
				(PT) edge [arc, bend left, below right] node {$p,t,\CDpPlaying$}
		(Playing)

				(Playing) edge [arc] node {$vh$} (Playing-H)
				(Playing-H) edge [arc, bend angle=20, bend right, below] node
		{$vh$} (BusyH)
				(BusyH) edge [arc, bend angle=20, bend left] node {$vh$}
		(H-Playing)
				(H-Playing) edge [arc, bend angle=20, bend left] node
		{$p,t,\CDpPlaying$} (Playing)
				(Paused) edge [arc, above] node {$vh$} (Paused-H)
				(Paused-H) edge [arc, above right] node {$vh$} (BusyH)
				(BusyH) edge [arc, bend angle=15, bend right, left] node {$vh$}
		(H-Paused)
				(H-Paused) edge [arc, below] node {$p,t,\CDpPaused$} (Paused)

				(Playing) edge [arc, bend angle=10, bend left, below left] node
		{$vh$} (Playing-Paused)
				(Playing-Paused) edge [arc, bend angle=10, bend left, right]
		node {$p,t,\CDpPaused$} (Paused)
				(Paused) edge [arc, bend angle=10, bend left, above right] node
		{$vh$} (Paused-Playing)
				(Paused-Playing) edge [arc, bend angle=10, bend left, left] node
		{$p,t,\CDpPlaying$} (Playing)

				(Playing) edge [arc, right] node {$vh$} (Playing-BUSYF)
				(Playing-BUSYF) edge [arc] node {$vh$} (BUSYF)
			;

			\def\dist{9}

			\node [place] at (\dist, 14) (Closed) {Closed}; 
			\node [place] at (\dist, 10) (Open) {Open};
			\node [transition] at (\dist+.75, 12) (Closed-Open) {load};
			\node [transition] at (\dist-.75, 12) (Open-Closed) {load};
			\node [transition] at (\dist, 16) (Closed-Closed)
		[label=above:{$[p=\faux]$}] {play};
			\node [transition] at (\dist+3, 12) (Open-Play-Closed)
		[label=right:{$[p=\faux]$}] {play};
			\node [transition] at (\dist+6, 12) (Closed-off) {off};
			\node [transition] at (\dist, 8.75) (Open-off) {off};
			\node [place] at (\dist, 7.5) (Off) {$S^F$};
			\node [place] at (\dist, 18.5) (Pout) {$\Pout$};

			\path
				(Closed) edge [arc, bend angle=10, bend left, right] node {$vh$}
		(Closed-Open)
				(Closed-Open) edge [arc, bend angle=10, bend left, right] node
		{$\faux,0,h$} (Open)
				(Open) edge [arc, bend angle=10, bend left, left] node {$vh$}
		(Open-Closed)
				(Open-Closed) edge [arc, bend angle=10, bend left, left] node
		{$\vrai,1,h$} (Closed)
				
				(Closed) edge [arc, bend left] node {$vh$} (Closed-Closed)
				(Closed-Closed) edge [arc, bend left] node {$vh$} (Closed)

				(Open) edge [arc, bend angle=20, bend right, below] node {$vh$}
		(Open-Play-Closed)
				(Open-Play-Closed) edge [arc, bend angle=20, bend right, right]
		node {$vh$} (Closed)

				(Closed) edge [arc, bend angle=20, bend left, above] node {$vh$}
		(Closed-off)
				(Closed-off) edge [arc, bend angle=20, bend left] node {$vh$}
		(Off)
				(Open) edge [arc] node {$vh$} (Open-off)
				(Open-off) edge [arc] node {$vh$} (Off)
			;

			\node [transition] at (4, 10) (Playing-Open) {load};
			\node [transition] at (4, 8) (Paused-Open) {load};
			\node [transition] at (5, 12) (BUSYF-Open) {load};
			
			\path
				(Playing) edge [arc, bend angle=15, bend right, below right]
		node [xshift=5, yshift=-5] {$vh$} (Playing-Open)
				(Playing-Open) edge [arc, below] node {$vh$} (Open)
				(Paused) edge [arc, below] node {$vh$} (Paused-Open)
				(Paused-Open) edge [arc, below] node {$vh$} (Open)
				(BUSYF) edge [arc] node {$vh$} (BUSYF-Open)
				(BUSYF-Open) edge [arc, below left] node {$vh$} (Open)
			;

		\end{tikzpicture}
		}
		\caption{Translation into a CPN (partial scheme)}
		\label{figure:CD:CPN}
	\end{subfigure}
	
}
	\caption{An example of a CD player~\cite{ZL10}}
        \label{figure:UMLCPN}
\end{figure}
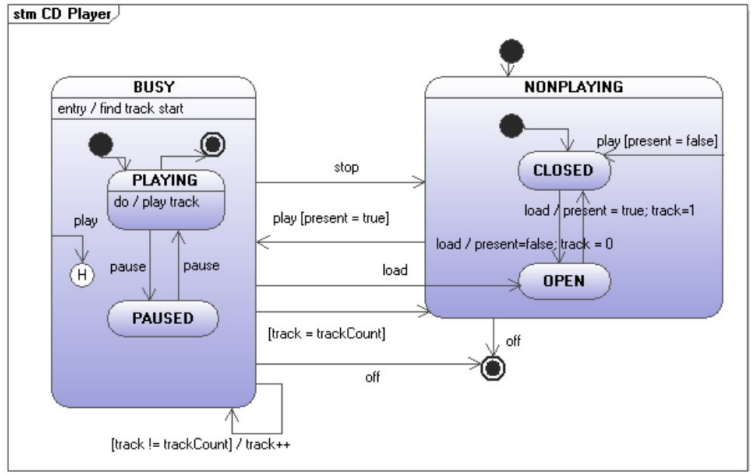
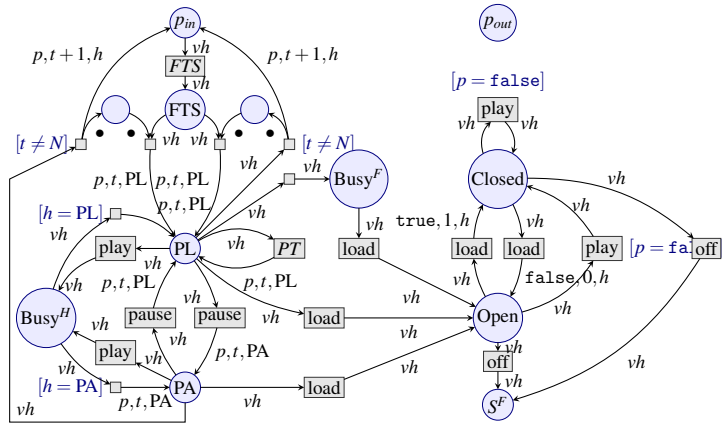

\paragraphe{UML State Machines}
%
We briefly introduce some of the SMD concepts with an example of a CD player (from~\cite{ZL10}) depicted in Figure~\ref{figure:CD:UML}.
This SMD is composed of two composite states (viz.\ BUSY and NONPLAYING), and each of them contains two simple states. 
BUSY has an entry behaviour, and PLAYING has a ``do'' behaviour.
Recall that entry/exit/do behaviours are behaviours 
performed when entering/exiting/being in a state, respectively.
Transitions may involve events, guards and behaviours,
where (global) variables may appear (e.g.\ \texttt{track}).
BUSY has a history pseudostate (depicted with 
``H'').

In our setting, we design UML state machines using the Eclipse Modeling Framework (EMF). 


\paragraphe{Coloured Petri Nets}
%
Petri nets are bipartite graphs with two kinds of nodes, viz.\ places 
and transitions. 
Directed arcs connect places to transitions 
and transitions to places. 
Coloured Petri nets (CPNs)~\cite{JK09} extend Petri nets with 
types.
Although this extension is mainly syntactic, it greatly increases the abstraction in complex Petri nets by making their representation more compact.
Figure~\ref{figure:CD:CPN} presents an example of a CPN that corresponds to the translation of the CD player of Figure~\ref{figure:CD:UML}, following the rules of~\cite{ACK12}.
In Figure~\ref{figure:CD:UML} state BUSY has a final state (represented in Figure~\ref{figure:CD:CPN} by Busy$^{F}$), a history state (represented by Busy$^{H}$) and an entry state (represented by transition/place FTS -- that stands for ``find track start'').
States PLAYING, PAUSED, CLOSED and OPEN are represented by PL, PA, Closed, and Open respectively.
State NONPLAYING is not represented because it has no final state, history pseudostate or behaviours.

\CPNTools{}~\cite{Westergaard13} is a powerful 
tool for modelling and verifying coloured Petri nets.
Whereas it features a user-friendly graphical interface, one can also design CPN models in an XML-based input syntax.
\ea{Is there a publicly available grammar for CPN Tools?}


\paragraphe{Acceleo}
%
Acceleo is a model-to-text transformation tool (its development started in 2006); its main purpose is to implement code generators.
An Acceleo program requires a metamodel and a model compliant with that metamodel, from which it generates source code.
Acceleo comes 
in the form of a user-friendly Eclipse plugin, hence easy to install, and features syntax highlighting tools, completion, etc.

The metamodel and the model are defined using EMF
, which makes Acceleo compatible with other tools based on EMF.
In particular, all models created from the ATL (Atlas Transformation Language) framework~\cite{ATL06} can be directly passed to Acceleo.

Acceleo is based on the notion of \emph{template}, that will implement the transformation rules.
Each rule in a template will map an element from the metamodel (and hence the model) to the text to be generated.
An advantage is that the structure of the Acceleo templates will directly reflect the structure of the generated code; in particular, the destination model is directly generated, with no need for postprocessing.

A main feature of Acceleo is that the generated text is mixed with Acceleo syntax.
Figure\,\ref{figure:ex:Acceleo} describes function $\mathit{SupEn}()$ (used in~\cite{ACK12}).
This function takes as parameters a state \texttt{s} and the global state machine, and writes to the output file the entry behaviour of~\texttt{s}.
This function mixes control structures to find the entry behaviour with the CPN code to be written to the output file.
Another example of the use of functions is given in Figure\,\ref{figure:template-function}.
The result \texttt{s.Name} is not written to the output file, but returned to the main template (that calls the function) as text.
After that, we can extract the result from the text.

Acceleo syntax offers some control structures such as \texttt{if} conditions, \texttt{for} loops, and \texttt{let} structures that allow to define variables, though with a limited scope.
However, to the best of our knowledge, Acceleo syntax features neither global variables, nor more complicated user-defined data structures (hash tables, lists, etc.), nor user-defined functions.

\begin{figure}[t]
	\begin{subfigure}[b]{0.48\textwidth}
\begin{lstlisting}
[template public SupEn1(s : State, pere : State, as : StateMachine)]
	[if (s.Entry().contains('true'))]
		[if (pere.Entry().contains('true'))]
			<arc id="ArcNRootSENS[s.Name/][s.Id/]"
           			orientation="PtoT"
           			order="1">
        			<posattr x="0.000000"
                 		y="0.000000"/>
     ...
[/template]
\end{lstlisting}
		\vspace{-1cm}
		\caption{Mixing Acceleo code with destination syntax}
		\label{figure:ex:Acceleo}
	\end{subfigure}
	\hfill
	\begin{subfigure}[b]{0.48\textwidth}
\begin{lstlisting}
[template public substates(s : State, as : StateMachine)]
  [if (s.isSimple = true)]
    [s.Name/]
  [else]
    [for (x : State | as.State)]
      [if (x.Parent = s.Name)]
        [substates(x, as)/]
      [/if]
    [/for]
  [/if]
[/template]
\end{lstlisting}
		\vspace{-1cm}
		\caption{Simulating functions with templates}
		\label{figure:template-function}
	\end{subfigure}
	\caption{Example of Acceleo code}
        \label{figure:acceleo}
\end{figure}

\section{Implementation of the Translation}

\subsection{Structure of the Translation}

We first briefly recall the structure of the translation of~\cite{ACK12}.
Each state (resp.\ behaviour) in the SMD is translated into a place (resp.\ transition) of the CPN.
Then, each transition in the SMD is translated to a set of new places and transitions in the CPN.
Various arcs are also added by the translation.
The translation of~\cite{ACK12} comes in the form of three algorithms translating the SMD states, transitions and history pseudostates, respectively.
These algorithms use several loops enumerating states, transitions, etc.
Furthermore, they use predefined functions (that retrieve, e.g.\ the substates of an SMD state, or the next exit behaviour to be executed after a given exit behaviour) and a mapping system that records in particular the translation  of a given SMD state into a CPN place.
Indeed, since several transitions in the SMD can have the same source and/or the same destination state, the translation process requires to be able to map (or store) easily the translation of a given SMD state into a CPN place.

\subsection{Toward Model-to-Text Transformation}

When implementing our translation, we had two main solutions: either implement a home-made translating tool (including lexers, parsers, and transformation functions based on an abstract syntax), or use model-to-model (M2M) transformation techniques.

M2M transformation techniques feature the following advantages: standard concepts and technologies, tools integrated in user-friendly frameworks (often Eclipse), and no need to write error-prone and time-consuming lexers, parsers, and printers.
The main drawback of M2M transformation techniques is that they usually require a metamodel for both the source model (UML state machines in our case) and the destination model (coloured Petri nets in our case).
Whereas the metamodel of UML state machines is provided by the OMG~\cite{UML241}, there is unfortunately no widely recognised metamodel for coloured Petri nets.
A solution could have been to translate UML state machines to non-coloured, classical Petri nets (``place/transition''), for which metamodels exist, but this solution is not satisfactory, since the translation would be much more cumbersome, and the resulting model much larger and less easy to understand. 
\ea{if we can find a justification (one day), that'd be good}

\subsection{Model-to-Text Transformation Using Acceleo}

The main advantage of model-to-text transformation techniques is that they do not require a destination metamodel.
As a consequence, Acceleo is adapted to our situation of a destination formalism without a widely recognised metamodel.
Hence, we considered our destination \CPNTools{} model (described in an XML-like syntax) as a \emph{text} document, hence suitable to be generated automatically using Acceleo.


\begin{figure}[t]
{\centering
	\includegraphics[width=.9\textwidth]{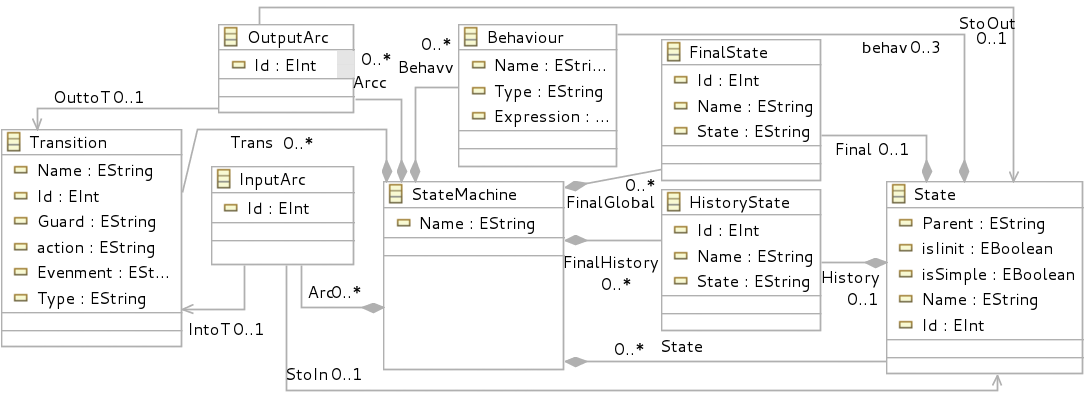}
	
}
\vspace{-.2cm}
\caption{Metamodel of UML state machines used in our translation}
\label{figure:metamodel}
\end{figure}

Our first step has been to set up a metamodel for UML state machines, given in Figure~\ref{figure:metamodel}.
We mostly reused the metamodel specified by the OMG~\cite{UML241}, with minor modifications:
indeed, we simplified this metamodel 
to the non-concurrent case, and performed a few minor changes (addition of some attributes that can be automatically derived from other attributes, and aim at easing the transformation).
In short, the system is represented using a global state machine (class \texttt{StateMachine}).
Then, each state machine is composed of states (classes \texttt{State} and \texttt{FinalState}), transitions (class \texttt{Transition}), behaviours (class \texttt{Behaviour}), pseudostates (class \texttt{HistoryState}) and arcs between states and transitions (classes \texttt{InputArc} and \texttt{OutputArc}).\ea{useful?!}

\subsection{Advantages and Limitations of Acceleo}

Our implementation of the translation mechanism of~\cite{ACK12} using Acceleo has been greatly facilitated by the user-friendly features of Acceleo: integration into the powerful Eclipse environment, and the use of the metamodel, avoiding us to deal with complex parsing mechanisms.
%
However, the translation also suffered from several drawbacks coming from the features (or lack of features) of Acceleo.

\paragraphe{Functions}
First, the lack of user-defined functions has been a major problem while implementing our translation.
Note that, even when changing the form of the algorithms of~\cite{ACK12}, the notion of function, in the sense of an operation depending on some input, would still be required by the translation.
However, the notion of template offered by Acceleo partially compensates the lack of function, at the price of a less clear code and lower performances.
Above all, Acceleo templates can only output text, which constrains us to rewrite our functions so that they output text, which also requires us to then parse their result.

Figure~\ref{figure:template-function} presents our translation of the $\mathit{substate}(s)$ function of~\cite{ACK12}, that returns the list of all substates (in a recursive manner) of an SMD state~$s$,
with a recursive template \texttt{substate}.
This solution is correct (in an algorithmic point of view) but certainly not optimal.
First, since there is no possibility to store the information computed previously (see below), one needs to iterate not only on the substates of a given state (since this information cannot be stored), but on all the SMD states.
This iteration is performed several times, leading to a cost exponential in the number of SMD states.
Second, this template returns a list of states in the form of a text; hence, checking that a state indeed belongs to this ``list" will require us to use the \texttt{substring} operation, hence again costing time.


\paragraphe{Global variables and data structures}
A second (and stronger) limitation of Acceleo is the impossibility to store information in the form of (global) variables or, more generally, of user-defined data structures such as arrays, lists, hash tables, etc.
This is a strong limitation, at least on an efficiency point of view: we were able to overcome this limitation, but at the cost of repeated recomputation of useful information, since it cannot be stored.




\subsection{Summary of our Experience}

We managed to implement using Acceleo the whole translation of SMDs into CPNs following the rules defined in~\cite{ACK12} (except the third algorithm of~\cite{ACK12}, i.e.\ the translation of history states, that was not implemented due to lack of time%
\footnote{%
	This work has been carried during Mohamed Mahdi Benmoussa's Master thesis at LIPN.
}).
Using the resulting tool \OurTool{}, we could successfully translate several simple examples of SMDs, thus allowing for verification using \CPNTools{}.
The Acceleo code of our translation is available on a dedicated Web page\footnote{%
	\url{http://lipn.univ-paris13.fr/~benmoussa/UML2CPN/}
}.
Nevertheless, we do not consider our overall experience of model transformation using Acceleo as entirely successful, for several reasons.

First, due to some exponential loops in our Acceleo code, we think that our tool may be very inefficient in the case of large SMD examples. 
This said, for the (small) examples we have translated using \OurTool{}, the translation time was always below 1\,second. 

Second, the lack of features (in particular global variables, functions and data structures) have been limitations during our implementation of the translation, may be not in a theoretical point of view, but in a practical point of view (leading to a larger and less clear code).

Third, Acceleo, as a code generator, is intrinsically dependent on the destination syntax.
Hence, a future change in the syntax of \CPNTools{} would lead us to considerably change our Acceleo templates.
One could argue that we could generate using Acceleo an abstract destination syntax instead of the \CPNTools{} concrete syntax;
but one of the advantages of model-to-text transformation is actually to directly generate the destination syntax without the need for an additional parsing tool. 
Similarly, we could have separated the destination syntax from the rest of the Acceleo code, by using templates that would either generate destination concrete code, or encode functions without any concrete code.
Whereas this would have been technically possible, that way of implementing the translation would have been in contradiction with a main feature of Acceleo which is to mix the templates with the destination code. 

For these reasons, we are not entirely convinced that our tool will be easy to maintain; in particular, a change in the \CPNTools{} syntax, or an improvement of the translation of~\cite{ACK12} (e.g.\ to add concurrency), may result in very large changes in our Acceleo code.


\section{Final Remarks and Perspectives}\label{section:conclusion}

We presented here a report on an automated translation of UML state machines to coloured Petri nets using Acceleo.
The resulting tool \OurTool{} allows to automatically generate a CPN that can then be analysed by \CPNTools{} and hence formally prove or disprove the original system safety.
Due to the problems we encountered when using Acceleo, we think that our tool may not be easily maintained. 

Whereas it is understandable that Acceleo was limited to only some features to keep it simple, some extensions would greatly increase its expressive power, and ease its usability and dissemination.
In particular, adding global variables, functions and user-defined data structures would certainly make easier the use of Acceleo even beyond its primary purpose (code generation).
Another problem we met is the lack of extensive documentation and examples online, maybe due to a still relatively small community.



\paragraphe{Perspectives} We were not entirely convinced by our experience of model-to-text translation using Acceleo,
and we will likely rewrite our translation mechanism using a home-made compiler-like tool, so as to be able to easily adapt it to future improvements of the translation (e.g.\ extension of the translation of~\cite{ACK12} to concurrent and/or timed state machines).

Also note that the resulting CPN may be simplified in some cases.
For instance, some places and transitions added by the translation may sometimes be unnecessary. 
These simplifications, that are beyond the scope of this paper, could help to speed up the automated verification of the resulting CPN.

Finally, although it goes beyond of the scope of this paper, 
a challenging future work is to formally prove the equivalence between the original SMD and the resulting CPN.
Of course, the problem is that the OMG does not formally define a formal semantics for SMDs.
However, we could reuse the operational semantics that we recently proposed for SMDs~\cite{LLACSWD13}, and define a trace equivalence taking into account active states, behaviours and events.

\paragraphe{Acknowledgement}
This work also benefited from Taieb Ben\,Niha's experience regarding a similar translation from UML activity diagrams to coloured Petri nets, during his internship at LIPN.
We would also like to thank Alessandro Tiso for his suggestions regarding Acceleo.\ea{useful?!}

\bibliographystyle{eptcs}
\bibliography{acceleo}

\end{document}